\documentclass[epj]{svjour}

\usepackage{subfigure}
\usepackage{amsmath,amssymb}
\usepackage{graphicx}
\usepackage{upgreek}

\newcommand{\eref}[1]{Eq.~(\ref{#1})}

\newcommand{\fref}[1]{Fig.~\ref{#1}}
\newcommand{\Sref}[1]{Section~\ref{#1}}

\newcommand{\ie}{i.e.}
\newcommand{\force}{\boldsymbol{F}}
\newcommand{\diffn}{\boldsymbol{D}}

\def \refl {\mathfrak{r}}
\def \trans {\mathfrak{t}}

\begin{document}

\title{Cavity cooling of atoms: Within and without a cavity}
\author{Andr\'e Xuereb\inst{1,2}\fnmsep\thanks{Corresponding author. Electronic address:\ andre.xuereb@aei.mpg.de}\and Peter Domokos\inst{3}\and Peter Horak\inst{4}\and Tim Freegarde\inst{2}}
\institute{Max-Planck-Institut f\"ur Gravitationsphysik (Albert-Einstein-Institut) and Leibniz Universit\"at Hannover, Callinstra\ss{}e 38, D-30167 Hannover, Germany \and School of Physics and Astronomy, University of Southampton, Southampton SO17~1BJ, United Kingdom \and Research Institute of Solid State Physics and Optics, H-1525 Budapest P.O. Box 49, Hungary \and Optoelectronics Research Centre, University of Southampton, Southampton SO17~1BJ, United Kingdom}
\date{Received: \today}

\titlerunning{Cavity cooling of atoms}

\abstract{We compare the efficiencies of two optical cooling schemes, where a single particle is either inside or outside an optical cavity, under experimentally-realisable conditions. We evaluate the cooling forces using the general solution of a transfer matrix method for a moving scatterer inside a general one-dimensional system composed of immobile optical elements. Assuming the same atomic saturation parameter, we find that the two cooling schemes provide cooling forces and equilibrium temperatures of comparable magnitude.}

\maketitle

\section{Introduction}
\begin{figure}[t]
 \centering
 \includegraphics{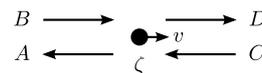}
 \caption{A scatterer, characterised by its polarisability $\zeta$ and velocity $v$, interacting with the four field modes that surround it.}
 \label{fig:Model}
\end{figure}
`Traditional' laser cooling schemes, such as those applied to the alkali atoms~\cite{Ashkin1978,Chu1998} {require} {a} closed set of two~\cite{Metcalf2003} or more~\cite{Dalibard1989} energy levels within which the atom cycles. These schemes rely on the spontaneous atomic decay from the excited to the ground state to carry atomic translational energy away from the system. The requirement for a closed set of levels {means that laser cooling has only been shown to be feasible for a small range of atomic species and for even fewer molecules}~\cite{Zeppenfeld2009,Shuman2010}.
\par
Cavity cooling~\cite{Horak1997} switches the energy decay process from {atomic spontaneous emission} to the decay of a cavity field. {Specifically}, the motion of an atom is coupled to a cavity field, itself driven either directly or through scattering from the atom~\cite{Domokos2002b}, and the decay of this field leads to damping of the atomic motion {by} a Sisyphus--like~\cite{Hechenblaikner1998} mechanism. This mechanism inherently relies solely on the dipole force and should therefore be applicable to any scatterer that is subject to the dipole force~\cite{Domokos2003}. This is the basis for recent investigations of cavity cooling of micromirrors~\cite{Thompson2008} {and dielectric} spheres~\cite{Barker2010}.\\
In a recent paper~\cite{Xuereb2010b} we used a scattering theory~\cite{Xuereb2009b} to examine the interaction between a particle and a `generalised interferometer', and proposed an alternative to cavity cooling whereby the particle to be cooled is not inside the cavity, but outside it. By physically separating the atom from the cavity, one can use cavities that are of much higher optical quality, or even solid-state, rather than macroscopic, cavities, thereby rendering the apparatus simpler and more amenable to miniaturisation. However, by placing the atom outside the cavity, the atom--cavity coupling is reduced significantly. The {important question} is whether, in practical implementations, these two effects compensate for one another in such a way as to render the cooling forces experienced by an atom outside a cavity similar to those it experiences inside.\\
To answer this question we will first {describe} the two models, in \Sref{sec:CMC} and \Sref{sec:ECCO}, respectively, using realistic parameters for state-of-the-art {optical devices}. Taking into account saturation effects, it is seen that the two models result in similar cooling forces and equilibrium temperatures. An examination of scaling properties of the force acting on the atom {in the two schemes} then follows in \Sref{sec:Scaling}, after which we conclude {in \Sref{sec:Conclusions}}.

\section{Comparison of cavity cooling schemes}\label{sec:ComparisonAtoms}
{
\subsection{Generic scattering model}
The two situations we describe will be discussed within the context of a scattering theory based on the transfer matrix method~\cite{Xuereb2009b}. This scattering model can substitute the usual cavity QED calculations for these systems. At each point in a one-dimensional space, and at each frequency, the electric field is described by two complex amplitudes, representing two waves moving in opposite directions, cf.~\fref{fig:Model}. The amplitudes $A$ and $B$ to the left of a generic scatterer, modelled through its polarisability $\zeta$, are related to $C$ and $D$ to its right by means of a $2\times 2$ matrix
\begin{equation}
\begin{pmatrix}
A\\
B
\end{pmatrix}=\begin{bmatrix}
1+i\zeta & i\zeta\\
-i\zeta & 1-i\zeta
\end{bmatrix}
\begin{pmatrix}
C\\
D
\end{pmatrix}\,.
\end{equation}
For a two-level atom in the low-saturation limit, we can write
\begin{equation}
\label{eq:ZetaAtom}
\zeta=-\frac{\sigma_\text{a}}{2S}\frac{\Gamma}{\Delta+i\Gamma}\,,
\end{equation}
where $\sigma_\text{a}$ is the on-resonance scattering cross-section of the atom, $S$ the mode area of the beam, $\Gamma$ the HWHM linewidth of the transition, and $\Delta$ the detuning of the pump beam from resonance with the transition. In the case of a mirror, $\zeta$ is related to its macroscopic properties via its reflectivity $\refl=i\zeta/(1-i\zeta)$ and its transmissivity $\trans=1+\refl=1/(1-i\zeta)$. Propagation of the electric field, having wavenumber $k$, over a distance $x$ in free space is represented by the matrix
\begin{equation}
\begin{bmatrix}
e^{ikx} & 0\\
0 & e^{-ikx}
\end{bmatrix}\,.
\end{equation}
In this model, complex optical systems can be built by multiplying the relevant matrices together, and the force acting on any single optical component is determined by the amplitudes of the fields interacting with it:
\begin{equation}
\label{eq:TMMForce}
\force_\text{full}=2\epsilon_0 S\bigl(\lvert A\rvert^2+\lvert B\rvert^2-\lvert C\rvert^2-\lvert D\rvert^2\bigr)\,.
\end{equation}
By accounting in the transfer matrix for the motion of the scatterer, following the process outlined in Ref.~\cite{Xuereb2009b}, the full velocity-dependent force can be calculated to first order in the scatterer velocity $v$ and the friction force $\force_1$ then extracted.
\par
A more rigorous model can be built that uses quantised fields rather than classical electromagnetic fields. In such a model, the transfer matrices operate on the respective annihilation operators, and a force operator can be defined in much the same way as the force in \eref{eq:TMMForce}. The two-time autocorrelation function of this force operator can then be used to obtain the momentum diffusion $\diffn$ acting on the scatterer~\cite{Xuereb2009b}. Finally, the fluctuation--dissipation theorem~\cite{Gordon1980} gives the equilibrium temperature $T$ that the motion of the scatterer will tend to:
\begin{equation}
k_\text{B}T=-\frac{\diffn}{\force_1/v}\,,
\end{equation}
which is only well-defined for a cooling force ($\force_1/v<0$), and where $k_\text{B}$ is the Boltzmann constant.}
\par
Further details of the calculations leading to the above expressions will not be presented in this paper. In the next two sections, the model is applied directly to investigate the nature of the cooling forces present in the two different configurations.

\subsection{Cavity mediated cooling: {Atom} inside the cavity}\label{sec:CMC}
\begin{figure}[t]
 \centering
 \subfigure[]{
  \includegraphics[scale=0.5]{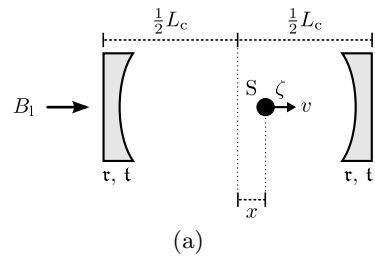}
 }\\
 \subfigure[]{
  \includegraphics[angle=-90,width=0.45\textwidth]{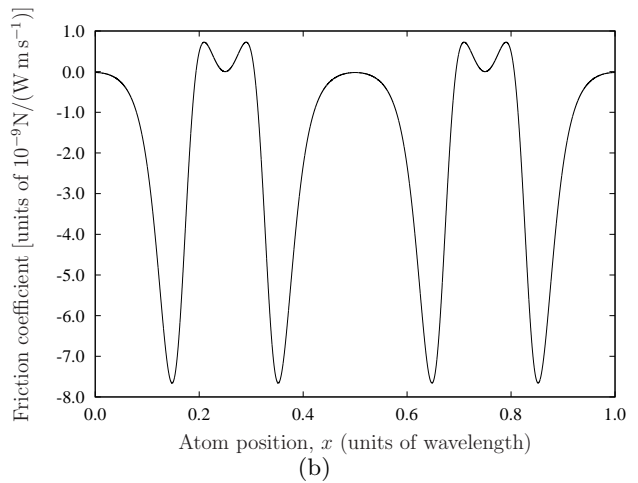}
 }
 \caption{(a)~{Model} of a {scatterer}, $\text{S}$, inside a symmetrical Fabry--P\'erot cavity of length $L_\text{c}$. The {cavity mirrors have} reflection and transmission coefficients, $\refl$ and $\trans$, and $\zeta$ is the polarisability of $\text{S}$. (b)~Friction coefficient, per unit input power, experienced by the scatterer at different positions in the cavity, for realistic parameters (see text for details).}
 \label{fig:CMC}
\end{figure}
Placing a scatterer---atom~\cite{Horak1997,Leibrandt2009,Koch2010}, micromirror~\cite{Bhattacharya2007a,Thompson2008}, or `point polarisable particle'~\cite{Domokos2003}---inside a cavity has long been pointed out to be a powerful means of cooling the translational motion of that scatterer. Cooling of atoms inside resonators has been observed: first~\cite{Leibrandt2009} as an increase in the storage time of atoms inside the cavity, and more directly in Ref.~\cite{Koch2010}. The layout of such an experiment is {shown} in \fref{fig:CMC}(a). For our purposes, we place the scatterer inside a symmetrical Fabry--P\'erot cavity of length $L_\text{c}$, which we pump from one side; the dominant field inside the cavity is a standing {wave} if the reflectivity of the mirrors, $\refl$, is sufficiently high. For a numerical example, we use the same cavity properties as Ref.~\cite{Mucke2010}: {finesse} $\mathcal{F}=56\,000$ {modelled by using mirrors with $\trans=\refl+1=1/(1+133.5i)$}, {cavity} length $495$\,$\upmu$m, and mode waist $30$\,$\upmu$m{; we use a wavelength $\lambda=780$\,nm. In contrast with Ref.~\cite{Mucke2010}, however}, our cavity is pumped along its axis.\\
We also take the scatterer to be a {two-level} atom, {with} the cavity {field detuned} $10\Gamma$ to the red of the atomic transition frequency. Thus, the polarisability of the atom is $\zeta=4.1\times 10^{-5}+4.1\times 10^{-6}i$. The maximum friction coefficient is found at a detuning of $-2.6$\,$\kappa$ from the cavity resonance. As expected~\cite{Domokos2003}, the optimal friction coefficient occurs for a negative {detuning, of the pump from} the \emph{bare} cavity resonance, but for a positive detuning from the {dressed atom--cavity} resonance.\par
The dependence of the friction force on the position of the {scatterer, scanned over a wavelength, is shown in \fref{fig:CMC}(b)}. {The presence of the cavity manifests itself primarily through a strong enhancement of both the friction coefficient $\force_1/v$ and the intracavity field intensity over their bare field values. The scattering model explored above is only valid in the limit of small saturation; e.g., when power broadening of the atom is negligible and the linear polarisability $\zeta$ in \eref{eq:ZetaAtom} applies. For the $^{85}$Rb D$_2$ transition, assuming that the beam is circularly polarised, the saturation intensity is $1.67$\,mW\,cm$^{-2}$~\cite{Steck2008}. In order to avoid saturation effects, we restrict the power input into the cavity to $2$\,pW; this equates to an intracavity intensity of $23$\,mW\,cm$^{-2}$ and hence a saturation parameter $s=0.14$; this is because $s$ is inversely proportional to the square of the detuning, $-10\Gamma$ in this case, of the pump beam from resonance. In turn, this input power yields a maximal friction coefficient of $-1.5\times 10^{-20}$\,N/(m\,s$^{-1})$, which corresponds to a $1/e$ velocity cooling time of $9$\,$\upmu$s for the same atom; averaging the friction force over a wavelength gives a cooling time of $37$\,$\upmu$s.}\par
In the low-saturation regime, the friction force and momentum diffusion both scale linearly with input power; the equilibrium temperature is therefore independent of the pump power. For the parameters used above, the equilibrium temperature predicted for a scatterer at the point of maximum friction is $56$\,$\upmu$K; averaging the friction coefficient, as well as the diffusion coefficient, over a wavelength, gives a higher equilibrium temperature of $220$\,$\upmu$K.

\subsection{External cavity cooling: {Atom} outside the cavity}\label{sec:ECCO}
\begin{figure}[t]
 \centering
 \subfigure[]{
  \includegraphics[scale=0.5]{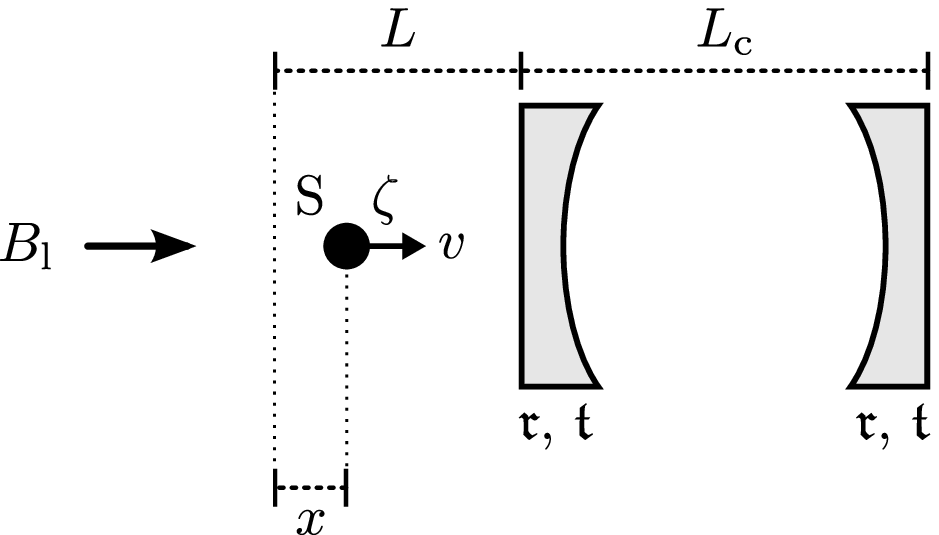}
 }\\
 \subfigure[]{
  \includegraphics[angle=-90,width=0.45\textwidth]{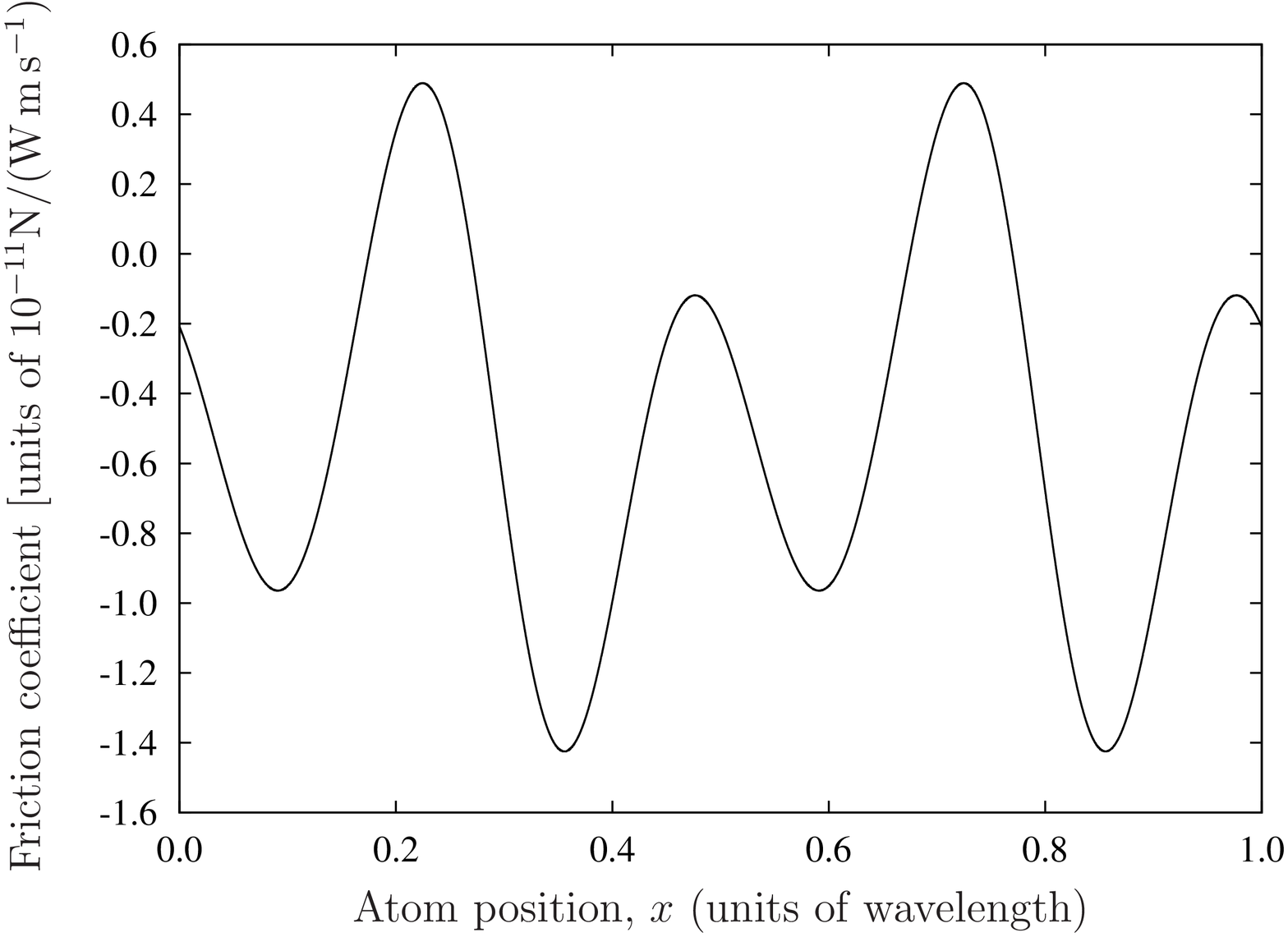}
 }
 \caption{{External cavity cooling. (a)~Model, similar to \fref{fig:CMC}, but with the atom at a distance $L-x$ outside the cavity.} (b)~Friction coefficient per unit input power experienced by the scatterer as $x$ is varied, for realistic parameters (see text for details). Note the change of scale, on the vertical axis, from \fref{fig:CMC}(b).}
 \label{fig:ECCO}
\end{figure}
More recently~\cite{Xuereb2010b} it was proposed that even with the scatterer \emph{outside} the cavity, the cavity's resonance can be exploited to enhance the optomechanical friction experienced by the scatterer over that in the standard optomechanical cooling setups~\cite{Braginsky1967,Metzger2004,Groblacher2009a}, which place the scatterer in front of a single mirror. It is the aim of this section to explore this cooling mechanism, using {experimental parameters similar to those in the previous section}, and compare it with the cavity mediated cooling mechanism discussed {there}.\par
Our mathematical model, \fref{fig:ECCO}, represents the cavity as a standard, symmetrical Fabry--P\'erot cavity. However, we emphasise that in principle what is required is simply an optical resonance: the cavity in the model can indeed be replaced by whispering gallery mode resonators~\cite{Schliesser2010} or even solid-state resonators. As a basis for numerical calculations{, and to enable direct comparison}, we model the same resonator as in the previous section. It is important to emphasise that the achievable quality factors of the resonators used for external cavity cooling can often be made larger than the ones in the previous section (see, e.g., Ref.~\cite{Rempe1992}) because {no} optical or mechanical access inside the resonator itself {is required}.
\par
The pump beam frequency is again taken to be detuned by $10\Gamma$ to the red of the atomic transition. By placing the atom outside the cavity, one is free to use high-numerical-aperture optics to produce a tighter focus than might be possible in a cavity with good optical and mechanical access. Having a tight focus strengthens the atom--field coupling because of the $1/w^2$ dependence of $\zeta$ on the beam waist $w$; whereas the friction force scales linearly with the input power, it also scales as $\zeta^2\sim 1/w^4$~\cite{Xuereb2010b}. Focussing the beam therefore increases the atom--field coupling more than the local intensity. {Thus, it is now assumed that the beam is focussed down to $1$\,$\upmu$m{, which} gives $\zeta=3.7\times 10^{-2}+3.7\times 10^{-3}i$.}\\
In order to make a fair comparison between the two cases, we choose to set the saturation parameter $s=0.14$, as in the previous section. The maximum achievable friction force coefficient is then {-$2.9\times 10^{-21}$\,N/(m\,s$^{-1}$) for $200$\,pW} of input power, {which compares} well with the previous result {and leads to a $1/e$ velocity cooling time of $50$\,$\upmu$s and an equilibrium temperature of $280$\,$\upmu$K. The magnitude of the force in this case results from the much smaller pumping beam mode waist, allowing the use of much higher powers and subsequently leading to a} stronger atom--field interaction. {W}ith this beam waist and finesse we would be restricted to input powers several orders of magnitude smaller if the atom were inside such a cavity.
\par
In {summary}, whereas the friction force inside a cavity is much stronger \emph{per unit input power and for the same beam waist}, the restrictions imposed on the magnitude of these quantities {when the atom is inside the cavity} reduce the maximally achievable friction force to a figure similar to {when it lies} outside the cavity.

\section{Scaling properties {of cavity} cooling forces}\label{sec:Scaling}
\subsection{Localisation}\label{sec:Localisation}
\begin{figure}[t]
 \centering
 \includegraphics[angle=-90,width=0.45\textwidth]{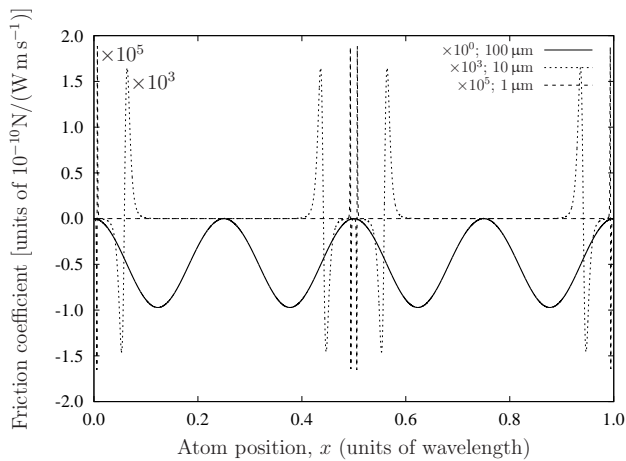}
 \caption{Spatial dependence of the friction force acting on an atom \emph{inside} a cavity with different mode waists but equal detuning from resonance, $10\Gamma$ to the red. The smaller the mode waist the stronger the friction force, by several orders of magnitude, but the more significant localisation becomes. (Parameters as in \Sref{sec:CMC} but with $\partial\zeta/\partial k=0$.)}
 \label{fig:CMC_Waists}
\end{figure}
\begin{figure}[t]
 \centering
 \includegraphics[angle=-90,width=0.45\textwidth]{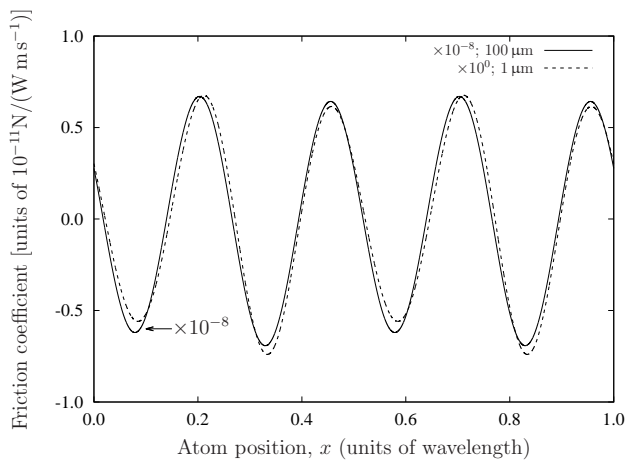}
 \caption{Spatial dependence of the friction force acting on an atom \emph{outside} a cavity, with different pumping field waists {but equal} detuning from resonance, $10\Gamma$ to the red. The friction force scales roughly as the inverse fourth power of the waist~\cite{Xuereb2009a}, but the length scale of the cooling and heating regions is unaffected. (Parameters as in \Sref{sec:ECCO} but with $\partial\zeta/\partial k=0$.)}
 \label{fig:ECCO_Waists}
\end{figure}
{The} broad nature of the spatial variations in the force shown in \fref{fig:CMC}(b) is {a consequence} of the small polarisability of the atom in such a cavity. {This is in sharp contrast to the case of large polarisability, achieved by an atom at a tight beam focus or by a micromirror, as shown} in \fref{fig:CMC_Waists}. A {scatterer of larger polarisability} would experience extremely narrow peaks, of spatial extent $\ll\lambda$, in the friction force if it lies inside a cavity but not outside it.\\
{Within the scattering model used in this paper, the atom--cavity coupling can be tuned by varying either the beam waist or the laser detuning from atomic resonance. Experimentally, however, atom--cavity coupling is rarely investigated close to resonance, in order to minimise the effects of atomic decoherence through spontaneous emission. In such cases, this coupling can be increased by operating a cavity with a small mode waist; this may in turn be detrimental to the performance of the system due to the strong sub-wavelength nature of the interaction, as explored in \fref{fig:CMC_Waists}.} The net {effect of having a smaller mode waist is} that this not only demands extremely good localisation but also tends to decrease the effective friction coefficient drastically{---by up to several orders of magnitude---}because of spatial averaging effects; this will be investigated in detail elsewhere.\par
In \Sref{sec:ECCO}, no mention was made of the average friction force acting on the scatterer; indeed this average computes to approximately zero for any case involving far-detuned atoms, or other particles with an approximately constant polarisability, outside cavities. This, then, also demands localisation of the atom {on a sub-wavelength scale}; whilst experimentally challenging this disadvantage is somewhat mitigated by the {easy} mechanical and optical access afforded by external cavity cooling schemes. Moreover, \fref{fig:ECCO_Waists} shows that spatial resonances are generally much wider here than in \fref{fig:CMC_Waists}: the polarisability of the atom can be varied over a very wide range without affecting the length scale of the cooling and heating regions.\\
In contrast with the atomic situation, if the scatterer is a micromirror mounted on a cantilever, localisation {does not present such a problem}, since such micromirrors naturally undergo small oscillations {and can be positioned with sub-nm accuracy}.

\subsection{Scaling with cavity {finesse and linewidth}}\label{sec:CavityLength}
{Cavity--mediated cooling mechanisms depend heavily on the physical properties of the cavity, namely its linewidth $\kappa$ and finesse $\mathcal{F}$. These parameters can be tuned independently by changing the length of the cavity and the reflectivity of its mirrors. This section briefly explores how the mechanisms considered scale with $\kappa$ and $\mathcal{F}$.
\par
Expressions for the force acting on an atom inside a good cavity are not simple to write down. Nevertheless, in the good-cavity limit one may obtain an analytic formulation for the limiting temperature~\cite{Horak1997}:
\begin{equation}
T=\frac{\hbar\kappa}{k_\text{B}}\,,
\end{equation}
\ie, making a cavity longer decreases the equilibrium temperature proportionally. This result can be understood by observing that whereas the diffusion constant depends only on the intensity inside the cavity ($\propto\mathcal{F}$), the friction force scales linearly with both the intensity and, if the intensity is kept constant, with the lifetime of the cavity field ($\propto 1/\kappa$). The friction force is therefore proportional to $\mathcal{F}/\kappa$, and the equilibrium temperature proportional to $\kappa$.}
\par
\begin{figure}[t]
 \centering
 \includegraphics[angle=-90,width=0.45\textwidth]{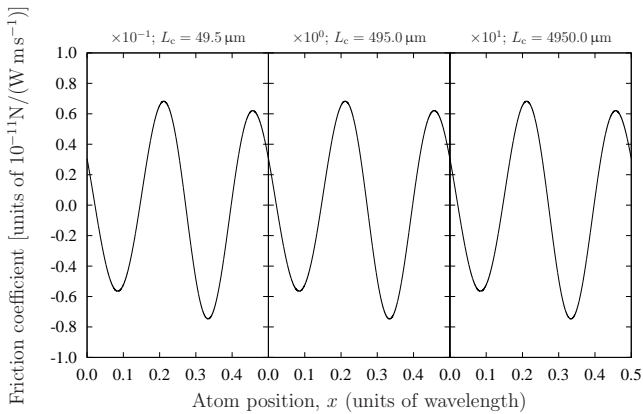}
 \caption{Spatial dependence of the friction force acting on an atom \emph{outside} a cavity, with detuning $10\Gamma$ to the red of resonance. Three different cavity lengths are shown; the friction force scales almost linearly with the cavity length. Note the different scaling factors and cavity lengths, given above each curve. (Parameters as in \Sref{sec:ECCO} but with $\partial\zeta/\partial k=0$.)}
 \label{fig:ECCO_Lengths}
\end{figure}
As is known from Ref.~\cite{Xuereb2010b}, the friction force ($\force_1$) acting on {an atom} outside a cavity scales {approximately} linearly with {both the length and the finesse} of the cavity. This is interpreted in terms of a `distance folding' mechanism: the lifetime of the light inside a cavity scales {inversely} with its {linewidth $\kappa\propto 1/\bigl(\mathcal{F}L_\text{C}\bigr)$} if all other parameters are kept fixed. One can see this {behaviour} reproduced in \fref{fig:ECCO_Lengths}, where the friction coefficient for on an atom outside each of three cavities having different lengths is shown. This mechanism loses its importance if {the atomic polarisability} is too large, whereby the system behaves more like two coupled cavities, or if the cavity is too long. {The momentum diffusion affecting the atom outside a cavity is essentially independent of the properties of a good cavity, since it depends on the local intensity surrounding the scatterer; putting these two results together, then, gives~\cite{Xuereb2010b}
\begin{equation}
T\approx 1.9\frac{\hbar\kappa}{k_\text{B}}\,,
\end{equation}
in the limiting case of small polarisability and at the point of maximum friction; \ie, the temperature scales in the same way as for an atom inside the cavity. The numerical factor in the preceding equation depends on $\zeta$ and is larger for $\zeta\sim 1$.}
\par
The cavity linewidth also affects the velocity `capture range' of the cooling mechanisms discussed in this text. In the case of cooling of atoms inside a cavity, it has been discussed in the literature~\cite{Horak2001,Domokos2002c} that the mechanism involved applies for atoms with a velocity $|v|<\kappa/k$. The theory described in this paper is correct only up to linear order in the velocity of the particle interacting with the cavity~\cite{Xuereb2009b}; within this framework, it does not seem possible to make predictions for the capture range. However, the external cavity cooling mechanism is expected to operate as described in the regime where the motion of the atom determines the slowest time-scale of the system; for the good-cavity limit, this would necessitate $|v|<\kappa/k$.

\section{Conclusions}\label{sec:Conclusions}
We have used the solution for the friction force and diffusion constant experienced by a polarisable scatterer interacting with a general 1D optical system to compare two different `cavity enhanced' cooling methods, where the scatterer is either inside or outside the cavity. It was found that, with state-of-the-art experimental parameters, the constraints imposed by the saturation of atomic transitions imply that the two mechanisms produce friction forces and lead to equilibrium temperatures that are of comparable magnitudes. The velocity capture range of both mechanisms scales linearly with the cavity linewidth, and is therefore expected to be similar in the two cases.
\par
It was also shown that for weakly polarisable scatterers (e.g., atoms in the low-saturation regime), positioning requirements are stronger for external cavity cooling due to a net-zero spatially averaged friction force. This conclusion is reversed for strongly polarisable scatterers (e.g., micromirrors), in which case the friction force inside a cavity varies over a much shorter length-scale than that outside.

\begin{acknowledgement}
This work was supported by the UK Engineering and Physical Sciences Research Council (EPSRC) grants\linebreak EP/E058949/1 and EP/E039839/1, and the European Science Foundation's EuroQUAM project {\em Cavity-Mediated Molecular Cooling}, and by the National Office for Research and Technology (ERC\_HU\_09 OPTOMECH) of Hungary.
\end{acknowledgement}

\bibliographystyle{epj}

\begin{thebibliography}{28}

\bibitem{Ashkin1978}
A.~Ashkin, Phys. Rev. Lett. \textbf{40}(12), 729 (1978)

\bibitem{Chu1998}
S.~Chu, Rev. Mod. Phys. \textbf{70}(3), 685 (1998)

\bibitem{Metcalf2003}
H.J. Metcalf, P.~van~der Straten, J. Opt. Soc. Am. B \textbf{20}(5), 887 (2003)

\bibitem{Dalibard1989}
J.~Dalibard, C.~Cohen-Tannoudji, J. Opt. Soc. Am. B \textbf{6}(11), 2023 (1989)

\bibitem{Zeppenfeld2009}
M.~Zeppenfeld, M.~Motsch, P.W.H. Pinkse, G.~Rempe, Phys. Rev. A \textbf{80}(4),
  041401 (2009)

\bibitem{Shuman2010}
E.S. Shuman, J.F. Barry, D.~DeMille, Nature \textbf{467}(7317), 820 (2010)

\bibitem{Horak1997}
P.~Horak, G.~Hechenblaikner, K.M. Gheri, H.~Stecher, H.~Ritsch, Phys. Rev.
  Lett. \textbf{79}(25), 4974 (1997)

\bibitem{Domokos2002b}
P.~Domokos, H.~Ritsch, Phys. Rev. Lett. \textbf{89}(25), 253003 (2002)

\bibitem{Hechenblaikner1998}
G.~Hechenblaikner, M.~Gangl, P.~Horak, H.~Ritsch, Phys. Rev. A \textbf{58}(4),
  3030 (1998)

\bibitem{Domokos2003}
P.~Domokos, H.~Ritsch, J. Opt. Soc. Am. B \textbf{20}(5), 1098 (2003)

\bibitem{Thompson2008}
J.D. Thompson, B.M. Zwickl, A.M. Jayich, F.~Marquardt, S.M. Girvin, J.G.E.
  Harris, Nature \textbf{452}(7183), 72 (2008)

\bibitem{Barker2010}
P.F. Barker, M.N. Shneider, Phys. Rev. A \textbf{81}(2), 023826 (2010)

\bibitem{Xuereb2010b}
A.~Xuereb, T.~Freegarde, P.~Horak, P.~Domokos, Phys. Rev. Lett.
  \textbf{105}(1), 013602 (2010)

\bibitem{Xuereb2009b}
A.~Xuereb, P.~Domokos, J.~Asb{\'{o}}th, P.~Horak, T.~Freegarde, Phys. Rev. A
  \textbf{79}(5), 053810 (2009)

\bibitem{Gordon1980}
J.P. Gordon, A.~Ashkin, Phys. Rev. A \textbf{21}(5), 1606 (1980)

\bibitem{Leibrandt2009}
D.R. Leibrandt, J.~Labaziewicz, V.~Vuleti{\'{c}}, I.L. Chuang, Phys. Rev. Lett.
  \textbf{103}(10), 103001 (2009)

\bibitem{Koch2010}
M.~Koch, C.~Sames, A.~Kubanek, M.~Apel, M.~Balbach, A.~Ourjoumtsev, P.W.H.
  Pinkse, G.~Rempe, Phys. Rev. Lett. \textbf{105}(17), 173003 (2010)

\bibitem{Bhattacharya2007a}
M.~Bhattacharya, P.~Meystre, Phys. Rev. Lett. \textbf{99}(7), 073601 (2007)

\bibitem{Mucke2010}
M.~M{\"{u}}cke, E.~Figueroa, J.~Bochmann, C.~Hahn, K.~Murr, S.~Ritter, C.J.
  Villas-Boas, G.~Rempe, Nature \textbf{465}(7299), 755 (2010)

\bibitem{Steck2008}
D.A. Steck (2008), {Rubidium 85 D Line Data},
  \texttt{http://steck.us/alkalidata/rubidium85numbers.pdf}

\bibitem{Braginsky1967}
V.B. Braginsky, A.B. Manukin, Soviet Physics JETP \textbf{25}(4), 653 (1967)

\bibitem{Metzger2004}
C.H. Metzger, K.~Karrai, Nature \textbf{432}(7020), 1002 (2004)

\bibitem{Groblacher2009a}
S.~Gr{\"{o}}blacher, K.~Hammerer, M.R. Vanner, M.~Aspelmeyer, Nature
  \textbf{460}(7256), 724 (2009)

\bibitem{Schliesser2010}
A.~Schliesser, T.J. Kippenberg, arXiv e-prints  (2010),
  \texttt{arXiv:1003.5922}

\bibitem{Rempe1992}
G.~Rempe, R.J. Thompson, H.J. Kimble, R.~Lalezari, Opt. Lett. \textbf{17}, 363
  (1992)

\bibitem{Xuereb2009a}
A.~Xuereb, P.~Horak, T.~Freegarde, Phys. Rev. A \textbf{80}(1), 013836 (2009)

\bibitem{Horak2001}
P.~Horak, H.~Ritsch, Phys. Rev. A \textbf{64}(3), 033422 (2001)

\bibitem{Domokos2002c}
P.~Domokos, T.~Salzburger, H.~Ritsch, Phys. Rev. A \textbf{66}(4), 043406
  (2002)

\end{thebibliography}

\end{document}